\documentclass[
preprint,
 amsmath,amssymb,
 aps,
prper,
floatfix,
]{revtex4-2}

\usepackage{graphicx}
\usepackage{dcolumn}
\usepackage{bm}
\usepackage{tabularx}

\usepackage{enumitem}

\usepackage{hyperref}
\hypersetup{colorlinks=true,urlcolor=blue,citecolor=blue,linkcolor=blue}
\usepackage{enumitem}          
\setlist{nosep}                 

\begin{document}

\title{Chained computerized adaptive testing for the Force Concept Inventory}

\author{Jun-ichiro Yasuda}
\affiliation{Center for the Studies of Higher Education, Nagoya University, Furo-cho, Chikusa-ku, Nagoya, Aichi, 464-8601, Japan}
\author{Michael M.~Hull}
\affiliation{University of Alaska Fairbanks, Department of Physics, 1930 Yukon Dr, Fairbanks, Alaska 99775, USA}
\author{Naohiro Mae}
\affiliation{Research Center for Nuclear Physics, Osaka University, Ibaraki, Osaka 567-0047, Japan}
\author{Kentaro Kojima}
\affiliation{Kyushu University, Faculty of Arts and Science, 744 Motooka, Nishi-ku, Fukuoka, 819-0395, Japan}

\date{\today}

\begin{abstract}
  Although conceptual assessment tests are commonly administered at the beginning and end of a semester, this pre-post approach has inherent limitations.  Specifically, education researchers and instructors have limited ability to observe the progression of student conceptual understanding throughout the course. Furthermore, instructors are limited in the usefulness of the feedback they can give to the students involved. To address these challenges, we propose an alternative approach that leverages computerized adaptive testing (CAT) and increasing the frequency of CAT-based assessments during the course, while reducing the test length per administration, thus keeping or decreasing the total number of test items administered throughout the course.  The feasibility of this idea depends on how far the test length per administration can be reduced without compromising the test accuracy and precision.  Specifically, the overall test length is desired to be shorter than when the full assessment is administered as a pretest and subsequent post-test.  To achieve this goal, we developed a CAT algorithm that we call \textit{Chain-CAT}.  This algorithm sequentially links the results of each CAT administration using collateral information.  We developed the Chain-CAT algorithm using the items of the Force Concept Inventory (FCI) and analyzed the efficiency by numerical simulations.  We found that collateral information significantly improved the test efficiency, and the overall test length could be shorter than the pre-post method.  Without constraints for item balancing and exposure control, simulation results indicated that the efficiency of Chain-CAT is comparable to that of the pre-post method even if the length of each CAT administration is only 5 items and the CAT is administered 9 times throughout the semester.  However, when these constraints are imposed, we found that the efficiency of the Chain-CAT with the test length of 5 items is lower than that of the pre-post method.  Based on this result and analysis of which items were administered, we suggest expanding the FCI item bank by creating additional items or incorporating items from other research-based assessments that have high discrimination parameters and exhibit sufficient variability in their difficulty parameters.
\end{abstract}

\maketitle

\section{\label{sec:Intro} Introduction}

When measuring pedagogical effectiveness, it is common practice to administer assessment tests before and after instruction.  After collecting the pre- and post-test scores, we calculate a statistic such as average normalized gain and analyze the average change in students' understanding due to instruction.  The results can then be compared with records from previous years or the reference values in the literature to determine the effectiveness of the instruction.  For example, the Force Concept Inventory (FCI) \cite{Hestenes1992} is an assessment test used to probe students' conceptual understanding of Newtonian mechanics.  The test has 30 items with five choices, and students typically take 20 to 30 minutes to complete the test.  Hake \cite{Hake1998} compared the average normalized gain of the FCI between interactive engagement courses and traditional courses and found that the interactive engagement method is more effective. There are many other recent studies administering the FCI or other assessment tests both before and after instruction and analyzing the effects of that instruction in terms of improvement of students' scores \cite{Pawl2020,Wilcox2020,Burkholder2020,Xiao2020,Reinhard2022,Wheatley2022,Morley2023,Christman2024}.

The pre-post method is a standard method to analyze the effectiveness of instruction, but there are limitations to the approach.  First, the pre- and post-test results provide only snapshots of students' understanding at the beginning and end of the course, limiting the ability to observe the progression of conceptual understanding throughout the duration of the course.  Second, students may see little benefit in their taking the assessment when the focus of the assessment is to reflect on the year's instruction and improve instructional practices for future students, with no feedback to the students who actually take the assessment.  It is therefore not surprising that post-test response rates are generally lower than pretest response rates (especially when students are asked to take the assessment outside of class), unless extra credit is offered to maintain student engagement and motivation \cite{Jariwala2016a,Nissen2018,VanDusen2021}.


To address these issues associated with the pre-post method, we propose an alternative approach which increases the frequency of the assessments during the course while decreasing the test length per administration, thus maintaining or decreasing the total number of test items during the course.  This idea is inspired by the microgenetic method \cite{Siegler1991,Kuhn1995,Sayre2009a,Heckler2010,Sayre2012,Brock2017,Brock2020,Brock2024}, which involves frequent quantitative and/or qualitative measurements.  The microgenetic method involves 1) sampling data at a frequency that is assumed to be high compared to the rate of change of the phenomenon of interest and 2) collecting data for the entire period of change \cite{Brock2020}, allowing teachers to monitor the progression of students' understanding.  Although these methods are generally implemented with small student cohorts, Sayre and Heckler \cite{Sayre2009a,Heckler2010} used the microgenetic approach in frequently administering simple multiple choice conceptual questions to a large number of students.

While the pre-post method aligns with diagnostic and summative assessment, our idea of frequent administration of short assessments can be used as a form of formative assessment.  In terms of research instruments, the method of survey administration chosen should depend upon the research questions being investigated.  In terms of enhancing instruction, our method opens the door to providing automated feedback according to each student's set of responses.  We expect that this will increase the usefulness of the survey for students and reduce their sense of burden in completing it.  

The feasibility of the approach we are proposing depends on how far the test length per administration can be reduced.  A reference value of the total test length during a course is 60 items, as this is the total test length if the full FCI is administered both as the pre- and post-test.  If the course involves 10 administrations of the shortened FCI, it is then desired that the test length per administration would be less than 6 items.

To reach this goal, we use computerized adaptive testing (CAT)  \cite{vanderlinden2010,Magis2017a}.  CAT is a practice in which a computer administers successive test items to match the current estimate of the student's proficiency (see below for details).  In one popular model of CAT, if a student answers an item correctly, the student will next need to answer a more difficult item.  On the other hand, if a student answers an item incorrectly, the student will answer an easier item next.  In this way, high (low) proficiency students do not need to answer items that are too easy (difficult) for them; thereby, the test length can be significantly shortened compared to standard test administration without compromising the accuracy and precision of the test. (Accuracy is the level of agreement between a measured value and a true value, and precision is the level of agreement between measured values obtained by replicate measurements on similar objects under specified conditions \cite{Joint2012}.)  Because of its efficiency, CAT is becoming widely used, for example, with the Graduate Record Exam (GRE)  \cite{Mills2000}, with SAT \cite{SAT2024}, with PISA  \cite{Frey2023}, and recently in physics/science education research \cite{Morphew2018,Samsudin2020,Linderman2021,Zakwandi2024,Morphew2024,Le2024}.

We utilize a CAT algorithm that takes advantage of collateral information \cite{vanderlinden1999,vanderlinden2010}.  Collateral information refers to relevant empirical data about respondents, such as age, grade level, or previous test scores.  In our study, we specifically used only the proficiency estimate from each respondent's previous test—that is, the most recent proficiency estimate carried over, as is commonly done (e.g., in Ref.~\cite{vanderlinden2010}).  This information can be used to select the first item in CAT, accelerating the convergence of the proficiency estimation, and hence improving test efficiency \cite{vanderlinden2010}.  Specifically, as we describe below, we use the proficiency estimate of each respondent in a test for selecting items and estimating respondent proficiency level in the next test for the respondent.  In contrast to one-time testing, these subsequent tests constitute multiple measurements over time (e.g., with intervals of about a week between them).  Since this CAT algorithm sequentially links the test result of each administration, we call this algorithm \textit{Chain-CAT}.  In CAT, the proficiency estimate is updated after each item and used to select the next item at a given point in time using a specified algorithm. In addition to this, Chain-CAT uses collateral information, employing the same algorithm with the proficiency estimate from the last question of the previous test (e.g., one week before) to initiate item selection in the current test.  Although the algorithm is the same, the term ``collateral information'' is introduced to emphasize the time gap between tests, during which learning can occur.

\begin{figure}[t]
  \begin{center}
  \includegraphics[width=14cm]{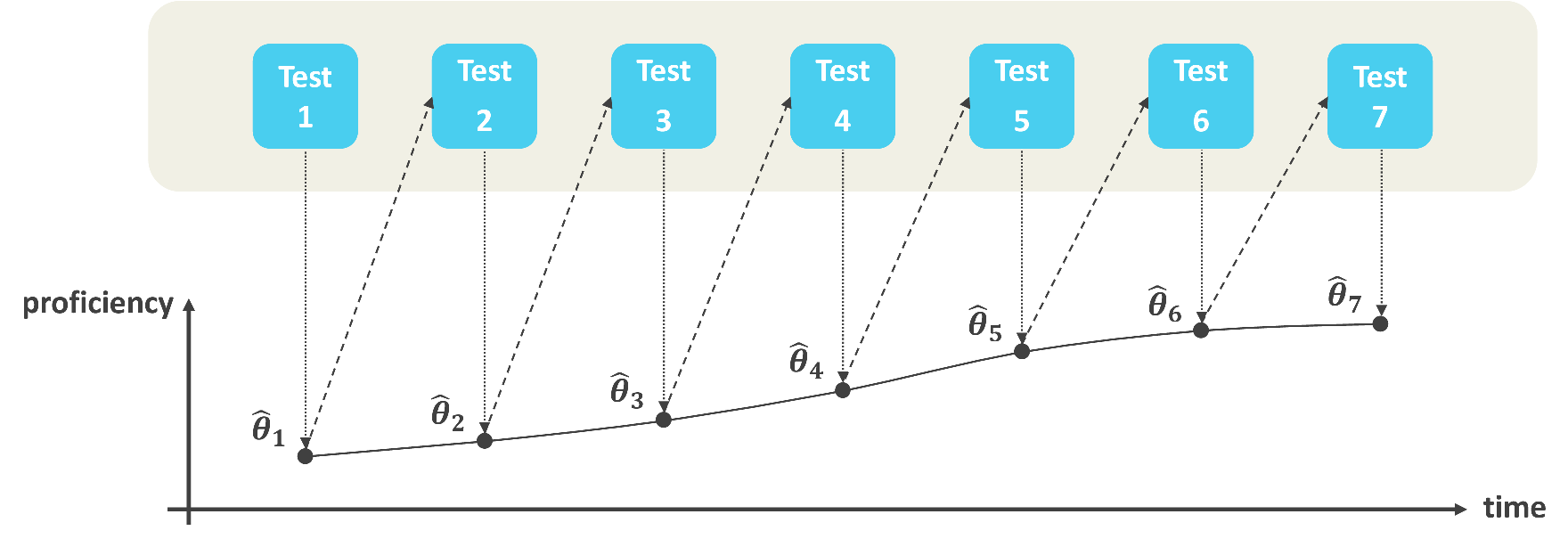} \\
  \end{center}
  \caption{\footnotesize Illustration of Chain-CAT algorithm for a student taking the tests for a course with seven administrations of the Chain-CAT.}
  \label{Fig1}
  \end{figure}

Figure~\ref{Fig1} illustrates the Chain-CAT algorithm in the case when a student takes the tests for a course with seven administrations of the Chain-CAT.  Each test is computerized adaptive, and each test length is relatively small (e.g., from 5 to 10 items).  After the student takes Test 1, his or her proficiency level is estimated (represented as $\hat{\theta}_1$ in the figure), and it is used as collateral information for the item selection and proficiency estimation at Test 2.  This procedure is continued sequentially from the next test to the final test (Test 7).

Several researchers have studied the development of a CAT-based formative assessment system.  Veldkamp \textit{et al.} \cite{Veldkamp2011} discussed the viability of a computerized adaptive learning system which responds to students based upon their test results.  Choi and McClenen \cite{Choi2020} developed and analyzed the validity and efficiency of an adaptive formative assessment system based on CAT with the time series proficiency estimation utilizing Bayesian networks.  Yang \textit{et al.} \cite{Yang2022} developed a formative assessment system based on CAT and learning memory cycle.  While there have been previous studies such as these, our goal is to construct an algorithm from our design of chaining collateral information as described above and to introduce its practical application in the context of physics education research.

The objectives of Heckler and Sayre \cite{Heckler2010} and our Chain-CAT are partially aligned, particularly in terms of tracking student understanding. However, the methods employed differ significantly. Heckler and Sayre \cite{Heckler2010} repeatedly administered individual questions to randomly selected groups of students and analyzed changes in the correct response rate of each question. In contrast, our approach employs a CAT-based system in which the test consists of multiple questions.  The selection of those questions varies depending on student's responses and the analysis is based on the proficiency estimates derived from those responses.

The objective of this study is to examine the feasibility of the Chain-CAT version of the FCI (Chain-CAT FCI).  Specifically, our research questions are: (i) How much does collateral information improve the test efficiency when it is used to sequentially link the results of each CAT administration? (ii) Is it possible to shorten the total test length of Chain-CAT FCI in a course to be less than that of the full FCI conducted as the pre- and post-test (60 items) without compromising test accuracy and precision?  Under what conditions is this possible?  (iii) Do the 30 FCI items comprise a sufficient item bank for the effective use of Chain-CAT? If not, what improvement is necessary? The focus of this paper is on the analysis of the feasibility of the Chain-CAT FCI using numerical simulations.  The development of feedback methods to deliver formative assessment will be the subject of future work.

The remainder of this paper is organized as follows.  In Sec.~\ref{sec:Method}, we present the design of the algorithm for Chain-CAT and the numerical simulation procedure analyzing its efficiency.  In Sec.~\ref{sec:Result}, we describe the simulation results and discuss the feasibility of the Chain-CAT FCI.  Finally, in Sec.~\ref{sec:Discussion}, we summarize this study and discuss limitations and future work. 

All our analyses were conducted using \textsc{R} \cite{RTeam2024}.  In addition to the basic package of \textsc{R}, the item parameters of the FCI were calibrated using the package \textsc{mirt} \cite{Chalmers2012} and the simulations of the Chain-CAT FCI were conducted using the package \textsc{catR} \cite{Magis2012,Magis2017b}.

\section{\label{sec:Method} Methodology}

\subsection{\label{subsec:IRT} Item response model and item bank}

We constructed the item bank of the Chain-CAT using the 30 FCI items.  The item parameters were calibrated based on the two-parameter logistic (2PL) model \cite{DeMars2012}.  In this model, the probability of a correct response from the $i$th respondent on item $j$ is given by
\begin{eqnarray}
  P_j (\theta_i) =\frac{1}{1 + \exp [-a_j(\theta_i - b_j)]},
  \label{eq:IRF}
\end{eqnarray}

In Eq.~(\ref{eq:IRF}), $\theta_i$ is the parameter representing the proficiency of the $i$th respondent.  The proficiency distribution in a reference population is standardized; namely, the estimated mean of $\theta_i$ is set to 0 and the estimated standard deviation of $\theta_i$ is set to 1.  In the equation, $b_j$ is the difficulty parameter, and $a_j$ is the discrimination parameter of item $j$.  The items with higher $a_j$ can better distinguish respondents who have different levels of proficiency.

We estimated the item parameters using the FCI survey data obtained in our previous study \cite{Yasuda2021,Yasuda2022}.  In that study, we administered the full-length paper-and-pencil FCI in class to 2882 Japanese university students from April 2015 to April 2018.  The examinees were students at the beginning of introductory physics courses at one public and four private universities, all of which are considered mid-ranked institutions in Japan.  From this dataset, we removed aberrant responses to be left with 2712 valid responses.  The result of the item parameter estimation is shown in Table~\ref{table3} in the appendix.  Although the standard error of the difficulty parameter for question 29 is relatively large, this item was rarely administered in the simulation due to its low discrimination. Hence, we expect that its inclusion has no significant effect on our results. 

The assumptions of unidimensionality, overall local independence, and goodness of fit were examined in our previous studies \cite{Yasuda2021,Yasuda2022}, supporting the validity of using the 2PL model in our analysis.  Specifically, we assessed the unidimensionality of the FCI using principal component analysis with a tetrachoric correlation matrix \cite{DeMars2012}, and evaluated the assumption of local independence using Yen's $Q_3$ statistic \cite{Yen1984}.  The results indicated that our FCI dataset has sufficient unidimensionality and local independence at the overall test level, although higher unidimensionality and local independence are desirable (see Sec.~\ref{subsec:limitation}).

We evaluated the goodness of fit of 1PL, 2PL, and 3PL IRT models to the response data with the standardized root mean square residual (SRMSR)  \cite{Maydeu2013}.  A value of SRMSR less than 0.05 is considered to indicate that the model is well-fitted.  The SRMSR values obtained were 0.079 for the 1PL model, 0.041 for the 2PL model, and 0.041 for the 3PL model.  These results suggest that both the 2PL and 3PL models fit the data well, whereas the 1PL model does not.

To select the most appropriate model, we used the Bayesian Information Criterion (BIC)  \cite{Schwarz1978}.  BIC increases if the deviance of the model from the data increases and if the number of parameters increases; thus, the model with the lowest BIC is the most preferable.  Based on our data set, the BIC was $9.22 \times 10^4$ for the 1PL model, $9.00 \times 10^4$ for the 2PL model, and $9.00 \times 10^4$ for the 3PL model.  These results indicate that the BIC values for the 2PL and 3PL models are comparable and both are lower than that of the 1PL model. 

Given the possibility that future studies may compare their results with those of the present study, it is important to reduce the cost of data collection. Since the recommended sample size for accurately estimating item parameters is approximately 500 for the 2PL model and around 2000 for the 3PL model  \cite{DeMars2012}, we selected the 2PL model for our analysis.

\subsection{\label{subsec:CAT} Basic computerized adaptive testing process}

The CAT process consists of four successive steps \cite{Magis2017a}: (i) initial step, (ii) test step, (iii) stopping step, and (iv) final step.  The basics for each step are as follows.

(i) \textit{Initial step}  The first item is selected and administered to a respondent.  The most used criterion to select the first item is the maximum Fisher information (MFI) criterion \cite{Magis2017a}.  The MFI criterion calls for selecting the most informative item (the item with the largest Fisher information) for the respondent based on the current estimate of proficiency. (Generally speaking, the ``most informative item'' is the one that will minimize the standard error of the proficiency estimate \cite{DeMars2012}.)  When nothing is known about the respondent, as is often the case when the first item is chosen for the first test, the Fisher information of the item is calculated using the mean proficiency value of the prior population, which is most often set to zero.  

(ii) \textit{Test step}  The proficiency of the respondent is estimated using the current set of item responses and the next item is selected to be administered.  For the proficiency estimation method, as in our previous study \cite{Yasuda2021,Yasuda2022}, we used the expected \textit{a posteriori} (EAP) method.  For the item selection criterion, we used the MFI criterion as in the initial step.  

(iii) \textit{Stopping step}  The test checks that a certain criterion has been met and the administration of the items ends.  We chose length to be the stopping criterion, such that CAT stops after a predetermined number of items have been administered, ranging from 1 to 30 items taken from the FCI.  

(iv) \textit{Final step} The final step involves the calculation of the final estimate of the respondent's proficiency level.  As in the test step, we chose the EAP method to estimate the proficiency level.

\subsection{\label{subsec:Chain} Design of the Chain-CAT algorithm}

In the Chain-CAT algorithm, as we described above, we linked the results of each CAT administration sequentially using collateral information. Specifically, each test incorporates proficiency estimates of previous tests as collateral information in the prior distribution for the proficiency estimation based on the Bayesian method.  This generally accelerates the convergence of the estimates in each CAT administration, thereby improving test efficiency.  

There are three stages where we can utilize collateral information: the initial step, the test step, and the final step of the testing process.  This process is completed within each ``Test'' in Fig.~\ref{Fig1}.  In what follows, we explain how we implemented collateral information for each stage.

(1) \textit{Initial step}  In this step, the first item is selected and administered to a respondent, as we described above.  For example, when the MFI criterion is used, the algorithm selects the item with the largest Fisher information for the current estimate of the respondent's proficiency.  At the beginning of the first test, when we know nothing about the respondents, the Fisher information of the candidate items is calculated using the mean proficiency value of the prior population.  This value is commonly set to be zero to have the scale be centered on respondents \cite{Magis2017a}, as we described above.  

At the beginning of each test after the first one, in the Chain-CAT algorithm, the Fisher information can be calculated utilizing the proficiency estimate of the former test for a given student as collateral information to improve the test efficiency for that student.  More generally, at the beginning of the $n$th $(n \geq 2 )$ test, the Fisher information can be calculated utilizing the $(n-1 )$th test proficiency estimate of a given student as collateral information.  

(2) \textit{Test step}  At this stage, CAT selects the next item based on the estimated proficiency level of the respondent at that point in the test.  We estimated the proficiency level using the EAP estimator, which is based on the Bayesian posterior distribution.  The posterior distribution for an $i$th respondent in turn is proportional to the product of the likelihood function and a prior distribution of the proficiency $g\left(\theta^i \right)$  \cite{Magis2017a}.  We used a model with a normal distribution, thereby we represent $g\left(\theta^i \right)\sim\mathcal{N}\left(\mu^i,\sigma^i \right)$, with a mean of $\mu^i$ and a standard deviation of $\sigma^i$. (Assuming a normal distribution is common in proficiency estimation using Bayesian item response modeling \cite{Magis2017a}.)  

On the first test, when we know nothing about the respondents beforehand, a common choice of the prior distribution is a normal distribution with $\mu^i=0$.  On the second test, we can utilize the first test proficiency estimate of an $i$th respondent, $\hat{\theta}_1^i$, as collateral information for the prior distribution.  Specifically, we chose the prior distribution as a normal distribution with $\mu^i=\hat{\theta}_1^i$ for the second test.  To generalize, we used the $(n-1)$th test proficiency estimate of an $i$th respondent, $\hat{\theta}_{n-1}^i$ as collateral information to determine the normal distribution such that $\mu^i=\hat{\theta}_{n-1}^i$ in the prior distribution for the $n$th test.  

The standard deviation of the prior distribution cannot be estimated in the method directly using the proficiency estimate.  In the simulation analysis (described in more detail below), we analyzed how the test efficiency depends on the value of $\sigma^i$ and searched for the optimal value which maximizes the accuracy and precision.    

(3)\textit{Final step}  We used collateral information also for the final proficiency estimation using the EAP method as just described for the test step.

\subsection{\label{subsec:valid} Improvement of the validity of Chain-CAT}

To increase the validity of the Chain-CAT FCI as a test of student understanding of Newtonian mechanics, one can impose constraints on how items are selected  \cite{Magis2017a}.  First, we implemented \textit{content balancing} for the Chain-CAT FCI as in our previous study \cite{YasudaHull2021}.  Content balancing ensures that the items in each test are not biased toward specific concepts (e.g., Newton's Third Law), and the same set of concepts assessed in the original test is covered in the CAT administration for each respondent.  To balance content in CAT, the percentage of items to be administered from each subgroup is defined in advance, for example, to be the same as what is found in the FCI itself.  Doing so ensures that items from each subgroup are administered.  The number of items and percentage of total items in each subgroup on the FCI is  \cite{YasudaHull2021}: Kinematics (5 items, 17\%), First law (8 items, 27\%), Second law (3 items, 10\%), Third law (4 items, 13\%), and Kinds of forces (10 items, 33\%).  

Various algorithms exist to control content balancing, but the \textsc{catR} package \cite{Magis2017a} allows the use only of the simplest option, the constrained content balancing method \cite{Kingsbury1989}.  The content balancing algorithm begins with the second item administered.  The steps of the algorithm selecting the second and subsequent items are  \cite{YasudaHull2021}:
\begin{enumerate}
  \item Before the administration of each item, compute the percentage of items that have already been administered from each subgroup.
  \item Target the optimal subgroup of items. Generally, this is the subgroup for which the gap between the observed relative proportion of administered items and the expected relative proportion is maximal. When multiple subgroups have not yet had any of their items administered, then one of those subgroups is chosen at random.
  \item Once the subgroup for the next item has been chosen in step 2, an item is selected from this subgroup to be the next item administered. As described above, the MFI criterion is used to choose the item from within the subgroup.
\end{enumerate}

For example, suppose that question 13 is selected for the first item.  This item is from the Kinds of forces subgroup.  Following the algorithm above, the second item will come from a subgroup that is chosen randomly from the remaining four subgroups. After an item from each subgroup is administered, the sixth item will be chosen from the Kinds of forces subgroup, since the gap between the observed relative proportion (20\%) and the expected relative proportion (33\%) is maximal.

Second, we controlled the \textit{item exposure} for the Chain-CAT FCI.  Generally, if the items in an item bank are very informative at the average ability level, they are selected more frequently than the other items \cite{Magis2017a}.  Although a given item does not appear more than once in a given test, the simple Chain-CAT algorithm can result in that item frequently appearing across the tests.  If students repeatedly answer certain items each week, they may memorize the content of the items they solved the previous week and answer them without rethinking, which may not reflect their actual proficiency.  Therefore, we implemented an algorithm that limits the number of times an item can be administered in a course.  Specifically, we used an algorithm whereby once an item is administered a specific number of times in a course (ranging from 3 to 9 items, depending on the test length), it is removed from the list of items that can be administered (described in more detail below).

\subsection{\label{subsec:sim} Numerical simulation procedure}

We conducted a numerical simulation to analyze the efficiency of the Chain-CAT and to search for the optimal settings.  The outline of the procedure is as follows.  (1) Assume a progression model of the true-value proficiency $\theta$ as a function of session $s$ (the unit is taken to be 1-2 weeks) in a course.  (2) Generate response data for a given true value of $\theta$ based on the item response model and calculate the estimates of $\theta$.  (3) Calculate the accuracy and precision by the root-mean-square error (RMSE) from the true and the estimated values of $\theta$ and compare it to a reference value.  In what follows, we explain the details of the above procedure.

\subsubsection{\label{subsubsec:prog} Progression models of the true proficiency level}

To analyze the efficiency of the Chain-CAT as generally as possible, we assumed various progression patterns of the true-value proficiency $\theta$.  Specifically, we assumed the following progression models for $\theta(s)$: a) a stationary model in which $\theta$ is constant across sessions, b) a linear model in which $\theta$ increases linearly across sessions, and c) a step model in which $\theta$ increases significantly only at a certain session.  Although it is uncommon in introductory physics courses for student scores on the FCI to decrease, we confirmed that the simulation results of the case when $\theta$ decreases linearly or decreases significantly only at a certain point are similar to the results of b) or c), respectively.  We supposed that most of the student's progressions are one of these patterns or a combination of them.  Moreover, we analyzed the test efficiency of the Chain-CAT in various proficiency ranges for each model.  For example, in the stationary model, we analyzed the efficiency in the middle $\theta$ students $(\theta=0)$ and the high $\theta$ students $(\theta=1.0)$. We confirmed that the low $\theta$ range $(\theta=-1.0)$ yields similar results to the high $\theta$ range $(\theta=1.0)$, so we do not show the result below.  We fixed the number of tests conducted throughout the simulated course to be 9 times.  This would be a reasonable number of tests to be administered in a course consisting of 10-15 weeks.  

\subsubsection{\label{subsubsec:gen} Generation of the response data}
We generated response data based upon a Monte Carlo simulation, which is commonly used in the development of CAT \cite{Thompson2011}.  Specifically, we generated the responses for Chain-CAT using the function \textit{simulateRespondents} of the \textsc{catR} package.  In the simulation, the response to a selected item is generated based on the calibrated item parameters of that item and the value of true proficiency at a particular session.  For example, in Eq.~(\ref{eq:IRF}), suppose that the probability of a correct response for an item is calculated to be 0.75 for a given true proficiency.  A random number is generated from a uniform distribution within a range of 0 to 1.  If the value is 0.75 or less, the generated response is coded as ``correct'' for that item.  If the value is greater than 0.75, then the generated response is ``incorrect.''  The EAP method is used to estimate the proficiency of the respondent, an item selection method (e.g., MFI criterion) is then used to choose the next item based upon that estimated proficiency, and the process repeats until reaching the predetermined test length.  Specifically, we varied the test length to be $L=$ 5, 10, and 15. Then, for each length, we calculated accuracy and precision.  

Note that when the test is administered nine times with test lengths of $L=5$, $10$, and $15$, the total number of items becomes 45, 90, and 135, respectively. In contrast, the pre-post method uses a total of 60 items. Therefore, only in the case of $L=5$ is the total number of items fewer than in the pre-post method. Our analysis in this paper includes results for $L=10$. We think such test lengths may be viable for instructors, even if the total number of questions exceeds 60. Limiting each session to 10 questions helps distribute the workload, potentially reducing the perceived burden. Moreover, even if the total number of questions is greater, having multiple small tests throughout the semester allows instructors to provide feedback as part of formative assessment.

To summarize, provided calibrated item parameters and a given true proficiency level, the \textit{simulateRespondents} function generates an entire set of correct/incorrect responses and calculates estimated proficiencies for a given test length of each Chain-CAT administration.  This is done for each session in a course.  In this analysis, we generated 10 000 response data based on the same true theta for a given session.

\subsubsection{\label{subsubsec:algoimp} Algorithm to improve the validity of Chain-CAT}

As we mentioned above, to improve the validity of Chain-CAT, we balanced the content and reduced the exposure of individual items. In the constrained content balancing method, items are selected so that the proportion of concept subgroups in an administered test is similar to that of the original test.  Let us assume the case of $L=5$, for simplicity.  The FCI has five subgroups ($N_\textrm{sub}=5$), as discussed above.  Thus, $L/N_\textrm{sub}$=1 item from each subgroup is administered in each test.  If we administer the test at $N_\textrm{S}$ (=9 in our simulation) sessions throughout the duration of the course, the total number of items to administer from each subgroup, $N_\textrm{tot}$, is thus given by $N_\textrm{S}$ times $L/N_\textrm{sub}$, which equals 9.   

In addition, to reduce exposure of individual items, we used an algorithm whereby once an item is administered a specific number of times, $N_\textrm{adm} (L)$, to a given respondent in a course, it is removed from the list of items that can be administered to that respondent.  The threshold $N_\textrm{adm}$ should be as small as possible; however, since there are only a limited number of items in the FCI and each subgroup, there is a lower limit to $N_\textrm{adm}$. If the number of times to administer a given item is set too low, there will be no items left to administer before the course has concluded. 

For example, the Second law subgroup has three items ($N_\textrm{cb}=3$). If $L=5$ and $N_\textrm{S}=9$, the total number of items to administer from each subgroup, $N_\textrm{tot}$, is equal to 9 from the above calculation, so each item in the subgroup should not be removed until it is administered 3 times  (=$N_\textrm{tot}$ divided by $N_\textrm{cb}$).  Therefore, in this case, we determine $N_\textrm{adm}$ to be 3.  For consistency, we use the same threshold in the other subgroups as well: for $L=5$ and $N_\textrm{S}=9$, a given FCI item is not administered more than three times.  

For the cases where the test length is longer than 5 items, the calculation of $N_\textrm{adm}$ is more complicated because the frequency of administering items is no longer the same for each subgroup.  We determined $N_\textrm{adm}$ by manually creating a table to lay out the order in which subgroups would be chosen for each subsequent item.  We then counted the number of items administered from each subgroup and derived that $N_\textrm{adm} (10)=5$, and $N_\textrm{adm} (15)=9$ are the minimum number of times administrated of a given item without resulting in subgroups becoming unrepresented in tests administered later in the semester. We confirmed these thresholds using numerical simulations.  Note that for $L=15$, $N_\textrm{adm}=9$ means that there is no limit because it equals the number of tests, $N_\textrm{S}=9$. That is, for $L=15$, a given FCI item can appear on each of the nine tests.

\subsubsection{Calculation of the accuracy and precision \label{subsubsec:calap}}

We represented the accuracy and precision by the bias and variance, respectively.  Then, we summarized these measures in terms of the mean-square error (MSE).  The MSE is a commonly used statistic for evaluating the accuracy and precision of parameter estimation in simulation studies. It is effective when the true proficiency parameters (theta) are known, particularly in the context of CAT \cite{Magis2017a}.  The MSE is defined by the following equation \cite{Bendat2012}, which equals the sum of the variance and the squared bias,
\begin{eqnarray}
  \mathrm{MSE}(\hat{\theta} )
  =E[(\hat{\theta}-\theta)^2 ]  
  = \textrm{Var} + B^2,
  \label{eq:mse}
  \end{eqnarray}

\noindent where $E(x)$ is the expected value of $x$.  In the equation, $\theta$ is the true proficiency given for each session and $\hat{\theta}$ is the estimated proficiency level.  The MSE was calculated as the average of the squared difference of $\theta$ and $\hat{\theta}$ using the above 10 000 response data for each session.  The bias $B$ is defined by, 
\begin{eqnarray}
  B(\hat{\theta} )=E(\hat{\theta})-\theta,
  \label{eq:bias}
  \end{eqnarray}

\noindent where $E(\hat{\theta})$ is represented by the average of the estimated proficiency level. The variance is given by,
\begin{eqnarray}
  \mathrm{Var}(\hat{\theta} ) = E [ (\hat{\theta} - E_{\hat{\theta}} )^2 ],
  \label{eq:var}
  \end{eqnarray}
\noindent where $E_{\hat{\theta}} =E (\hat{\theta} )$.  With the above definitions, we show our simulation results with the square root of the above statistics, namely, the root-mean-square error (RMSE) defined by $\textrm{RMSE}=\sqrt{\textrm{MSE}}$ and the standard deviation $\textrm{SD}=\sqrt{\textrm{Var}}$.

Our previous study \cite{Yasuda2021, Yasuda2022} indicated that, within the 30 items of the FCI, longer test lengths led to smaller RMSE.  Although generally, the higher the accuracy, the better, our approach prioritizes features such as test time and application to formative assessment, and so we look for a degree of accuracy that is comparable to that of the paper-based pre-post method. Specifically, we calculated the RMSE for the Chain-CAT and compared it to the RMSE obtained using the conventional pre-post method with the FCI. In the latter approach, all 30 FCI items were administered in both the pretest and post-test, and student proficiency was estimated using item response theory based on the same 2PL model described above, without collateral information. The RMSE from the pre-post method served as a reference value for evaluating the efficiency of the Chain-CAT.

When comparing the accuracy and precision of the Chain-CAT with the pre-post method, we use two different types of averages for the RMSE.  The first one is the session-averaged RMSE, which is defined by
\begin{eqnarray}
  \textrm{session.av.RMSE}
  =\frac{\sum_{s=s_\textrm{i}}^{s_\textrm{f}}\textrm{RMSE}(s)}{N_\textrm{S}}, 
  \label{eq:sessionav}
\end{eqnarray}

\noindent where $s_\textrm{i}$ is the initial session, $s_\textrm{f}$ is the final session, and $N_\textrm{S}$ is the number of the sessions (number of tests, as above).  The second type of average RMSE we calculated is the overall average of the RMSE, which is calculated by taking the average of the session-averaged RMSE for all progression models that we assumed in the simulations:
\begin{eqnarray}
  \textrm{all.av.RMSE}
  =\frac{\sum_{m=1}^{N_\textrm{M}}\textrm{session.av.RMSE}(m)}{N_\textrm{M}}, 
  \label{eq:allav}
\end{eqnarray}

\noindent where $m$ is the index of the progression model and $N_\textrm{M}$ is the number of the progression models we considered.  

\newpage

\section{\label{sec:Result} Result}

\subsection{\label{subsec:typ} Typical numerical simulation result}

\begin{figure*}[t]
  \begin{center}
  \includegraphics[width=13cm]{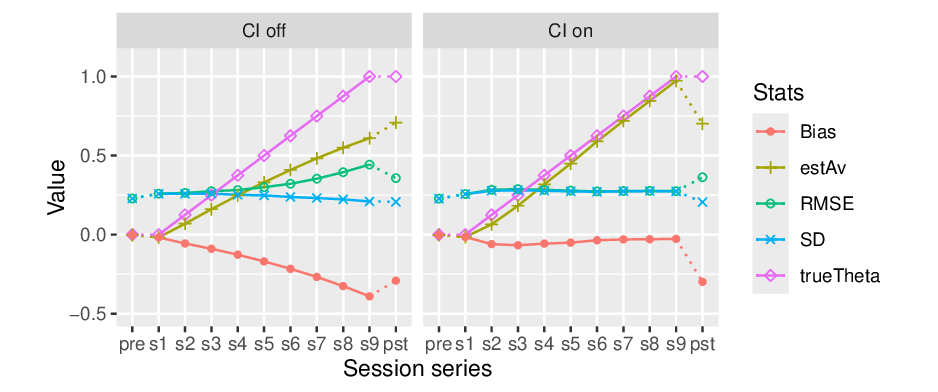} \\
  \end{center}
  \caption{\footnotesize Typical numerical simulation results for a linear progression model. The test length of each administration is set to $L=10$.  The standard deviation of the prior distribution was fixed at $\sigma^i=0.5$.  (a) Left: without collateral information (basic CAT). (b) Right: with collateral information (Chain-CAT).  The values of ``pre'' and ``pst'' show the results based on the pre-post method where the full FCI (30 items) is used in the proficiency estimation without collateral information.}
  \label{Fig2}
  \end{figure*}

Figure~\ref{Fig2} shows the simulation results assuming a linear model for the true proficiency progression $(0 \leq  \theta \leq  1)$.  The test length of each administration was set to $L=10$.  The standard deviation of the prior distribution was fixed at $\sigma^i=0.5$.  The left figure of Fig.~\ref{Fig2} is a result without collateral information (basic CAT) and the right figure is a result with collateral information.  The graphs include the simulation results for nine-sessions from $s_1$ to $s_9$, where the true proficiency level increases from $\theta(s_1)=0$ to $\theta(s_9)=1.0$ (see the purple lines in the graphs).  For each session, we generated 10 000 responses based on the same true theta at a given session and estimated the proficiency level for each response, then we calculated the average and the standard deviation of the estimated theta (the yellow and blue lines).  The standard deviation for the estimates of the $\theta$ mean is about 0.0025.  The bias was calculated from the difference between the true theta and the average of the estimated theta (the red lines), then the RMSE was calculated using Eq.~(\ref{eq:mse}) (the green lines).   In Fig.~\ref{Fig2} (b), the standard deviation is much larger than the bias, so the SD values and the RMSE values are very close.  In the graphs, the reference values are shown at $s=\textrm{pre}$ and $s=\textrm{pst}$ on the horizontal axis, where the statistics were calculated using the true proficiency levels at $s=s_1$ and $s=s_9$ and administering the whole test items (30 FCI items) in both cases without collateral information.  

From Fig.~\ref{Fig2}, we can observe that the collateral information significantly improved the accuracy.  With collateral information, the Fisher information is calculated using the proficiency level of the previous test, which prevents the bias from becoming larger when the true proficiency level is away from $\theta=0$.   Without the collateral information, the Fisher information is always calculated assuming the true proficiency level equals zero, therefore the deviation between the true theta and the estimated theta becomes large when the true proficiency level is away from $\theta=0$.  Even if we use the 30 FCI items, the bias is quite large when $\theta=1.0$, as shown at $s=\textrm{pst}$ in Fig.~\ref{Fig2}.

\subsection{\label{subsec:overall} Comparing overall trends}

As we mentioned above, the standard deviation of the prior distribution $\sigma^i$ is a free parameter.  We found that if $\sigma^i$ is small, the variance in Eq.~(\ref{eq:var}) tends to be small, but the bias in Eq.~(\ref{eq:bias}) tends to be large when the proficiency progression is large.  If $\sigma^i$ is large, the variance tends to be large, but the bias tends to be relatively small when the proficiency progression is large. Moreover, we found that the rank order of the overall average RMSE magnitude depends on the value of $\sigma^i$. For example, when $\sigma^i$ is set to 1 for all methods, the pre-post method shows the highest accuracy (lowest RMSE), whereas when $\sigma^i$ is set to 0.3, the Chain-CAT with collateral information shows the highest accuracy, and so on. We were concerned that comparing results at a fixed $\sigma^i$ could introduce arbitrariness, depending on the analyst's choice of the common $\sigma^i$ value. Therefore, we calculated the overall average across various $\sigma^i$ values and based our comparison on the $\sigma^i$ that yielded the lowest overall average RMSE. This approach reflects practical considerations for implementing the test in classroom settings, where $\sigma^i$ should be determined before administration without knowing the amount of proficiency progression in advance.  In this case, it is reasonable to choose $\sigma^i$ that yielded the lowest overall average in each method.

Specifically, for a given value of $\sigma^i$ (=0.1, 0.2, …, 1.0), we conducted simulations and calculated the session-averaged RMSE in Eq.~(\ref{eq:sessionav}) from $s=s_1$ to $s=s_9$. For the pre-post method, the session-averaged RMSE was calculated for $s=\textrm{pre}$ and $s=\textrm{pst}$. Then we calculated the overall average in Eq.~(\ref{eq:allav}) for all progression models that we assumed in the simulations.  After that, we chose the value of $\sigma^i$ which minimizes the overall average of the RMSE.  We obtained the following values for each case, namely, $\sigma^i=0.4$ for the case without collateral information, $\sigma^i=0.3$ for the case with collateral information, and $\sigma^i=0.6$ for the pre-post method.

\begin{table}[t]
  \caption{The session-averaged RMSE for each progression model ($L=10$)}
  \label{table1}
  \begin{tabular}{lrrr}
  \hline 
  \hline
  Progression model \hspace{1.6em} & \hspace{1.6em} CI off & \hspace{1.6em} CI on & \hspace{1.6em} Pre-post \\
  \hline
Constant ($\theta=0$)	& 0.210	& 0.186	& 0.252 \\
Constant ($\theta=1$)	& 0.531	& 0.289	& 0.322 \\
Linear ($0\leq \theta\leq 1$)	& 0.332	& 0.233	& 0.287 \\
Linear ($-0.5\leq \theta\leq 0.5$)	& 0.248	& 0.248	& 0.260 \\
Linear ($-1.0\leq \theta\leq 0$)	& 0.347	& 0.322	& 0.293 \\
Step ($0\leq \theta\leq 0.5$)	& 0.262	& 0.214	& 0.256 \\
Step ($-0.5\leq \theta\leq 0$)	& 0.250	& 0.250	& 0.256 \\
\hline
All model average	& 0.311	& 0.249	& 0.275 \\

  \hline \hline
  \end{tabular}
  \end{table}

After fixing the value of $\sigma^i$ which is most likely to give the highest efficiency for each case, we calculated the session-averaged RMSE.  Table~\ref{table1} illustrates the effect of collateral information within each progression model for the case of $L=10$. For each model, we compare the effectiveness of different methods: without CI, with CI, and the pre-post approach. To isolate and better understand the impact of collateral information, we did not incorporate content balancing or item exposure control in this table.  In Table~\ref{table1}, we can observe that the session-averaged RMSE in the case with collateral information is smaller than or comparable to that in the case without collateral information for all progression models.  Moreover, the session-averaged RMSE in the case with collateral information is smaller than that of the pre-post case for most of the progression models.  Especially, the case with collateral information is much more efficient when the absolute value of $\theta$ at final $s$ is large as in the Constant model ($\theta=$1) and the Linear model ($0\leq \theta \leq 1$). On the other hand, the case without collateral information has a comparable efficiency in the Linear model ($-0.5\leq \theta \leq 0$.5) and the Step model ($-0.5\leq \theta \leq 0$), where the absolute value of $\theta$ varies near zero and the absolute value of $\theta$ at final $s$ is small. One of the causes of this could be that the case without collateral information always assumes the mean value of prior distribution $\mu^i=0$.  Finally, the pre-post case is more efficient when the absolute value of $\theta$ at initial $s$ is large and at final $s$ is close to 0 as in the Linear model ($-1.0\leq \theta \leq 0$).  This is because for CAT with short test length, such as $L=10$, if the absolute value of $\theta$ at $s=s_1$   is large, the RMSE at $s=s_1$ is much larger than that of the pretest where $L=30$.

\subsection{\label{subsec:lci} Length dependence, content balancing, and item exposure control}

\begin{table}[t]
  \caption{The overall average RMSE for each test length with/without content balancing and item exposure control. The overall average RMSE is calculated by taking the average of the session-averaged RMSEs across all progression models assumed in the simulations, as defined in Eq.~(\ref{eq:allav}). These results are for the case with collateral information.}
  \label{table2}
  \begin{tabular}{lrrr}
  \hline 
  \hline
  Test lengh \hspace{2.5em} & \hspace{2.5em} 5 & \hspace{2.5em} 10 & \hspace{2.5em} 15 \\
  \hline
  CB and IE off & 0.277 & 0.249 &  0.233 \\
  CB and IE on  & 0.317 & 0.269 &  0.233 \\
  \hline \hline
  \end{tabular}
  \end{table}

Table~\ref{table2}  shows the dependence of the overall average RMSE of Chain-CAT (with collateral information) on the test length of each administration.  The overall average RMSE is calculated by taking the average of the session-averaged RMSEs across all progression models assumed in the simulations, as defined in Eq.~(\ref{eq:allav}).  The standard deviation of the prior distribution was fixed at the optimal value, $\sigma^i=0.3$.  Table~\ref{table2} also includes a comparison of the cases with/without content balancing and item exposure control.  In Table~\ref{table2}, we can see that the overall average RMSE increases as the test length decreases.  However, the overall averaged RMSE of Chain-CAT is comparable to that of the pre-post method  even when the test length is $L=5$ (compare the value of 0.275 in Table~\ref{table1} to that of 0.277 in Table~\ref{table2}).  This means that if we administer the Chain-CAT nine times in a course, by using the collateral information, we could reduce the total test length in a course to 45 items from the pre-post method (60 items) without compromising the test accuracy and precision if we are not concerned about improving the validity of the test.  On the other hand, by implementing content balancing and item exposure control, the overall average RMSE is increased from 0.277 to 0.317 when $L=5$, an increase of about 10\%.  This is because when the choice of administration is limited, the items that result in lower accuracy and precision are administered.  When $L=10$, the overall average RMSE with content balancing and item exposure control is smaller than that of the pre-post method (compare the value of 0.275 in Table~\ref{table1} to the value of 0.269 in Table~\ref{table2}).  In this case, the total test length is 90 items which is longer than that of the pre-post method, but the accuracy and precision are increased.  This improvement is partly attributable to the assessment using 50\% more items.

As noted above, applying content balancing or item exposure control limits the flexibility of item selection, resulting in the administration of items that reduce accuracy and precision. The findings above indicate that the current 30-item FCI is insufficient for implementing Chain-CAT effectively, and that expanding the item bank—either by developing new items or incorporating items from other concept inventories—is necessary. In this context, the results shown in Table~\ref{table1} and Table~\ref{table2}, obtained without content balancing or item exposure controls, offer valuable insights. They can be interpreted as representing a future scenario in which the FCI item bank has been expanded to include a sufficient number of replacement items with similar difficulty and discrimination characteristics. In other words, from these findings, we expect that such a level of measurement accuracy could be attainable if the FCI item bank is expanded, even with content balancing and item exposure controlled.
 
\begin{figure*}[t]
  \begin{minipage}[b]{0.45\linewidth}
    \centering
    \includegraphics[keepaspectratio, scale=0.65]{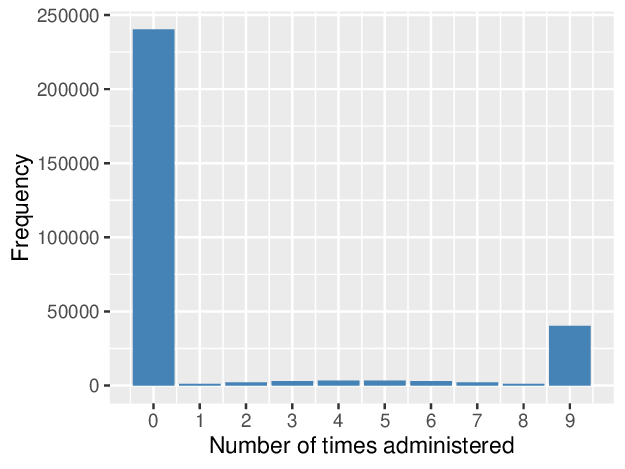}
  \end{minipage}
  \begin{minipage}[b]{0.45\linewidth}
    \centering
    \includegraphics[keepaspectratio, scale=0.65]{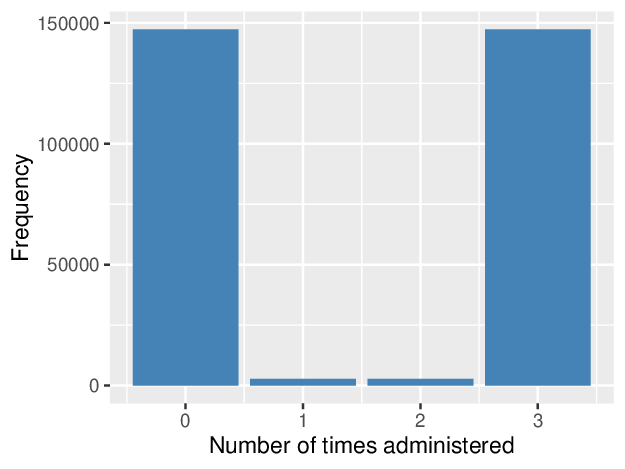}
  \end{minipage}
  \caption{\footnotesize Histograms of the number of times an item was administered in the nine tests when $L=5$ in the Linear model ($0\leq  \theta \leq  1$).  If an item appears in all nine tests, its frequency contributes one count to the bin labeled ``9'' in the histogram representing the number of times administrated.  The left (right) graph shows the result without (with) content balancing and item exposure control.  The sum of the products of each $x$ value (number of times administrated) and its corresponding $y$ value (frequency) equals the total number of item administrations in the simulation, which is 450 000.}
  \label{Fig3}
  \end{figure*}

Figure \ref{Fig3} shows histograms of how many times an item was administered in the nine tests when $L=5$ in the Linear model ($0\leq\theta\leq1$).  The horizontal axis shows the number of times an item was administered, and the vertical axis shows the frequency.  If an item appears in all nine tests, its frequency contributes one count to the bin labeled ``9'' in the histogram representing the number of times administrated. The left (right) graph shows the result without (with) content balancing and item exposure control.   Recall that our simulation had 10 000 simulees take nine tests.  In the case of $L=5$, this totals 450 000 items administered in the simulation.  In the histogram, the sum of the products of each $x$ value (number of times administrated) and its corresponding $y$ value (frequency) equals the total number of item administrations in the simulation, which is 450 000.  If we count any time that any of the 30 FCI items showed up in all nine of a simulee's tests, we find that occurs many times when we do not employ content balancing and item exposure control (right-most bar of the left graph in Fig.~\ref{Fig3}).  We can also see that there are even more cases where an item is never administered (left-most bar of the left graph in Fig.~\ref{Fig3}) in that semester.  

\begin{figure*}[t]
  \begin{center}
  \includegraphics[width=13cm]{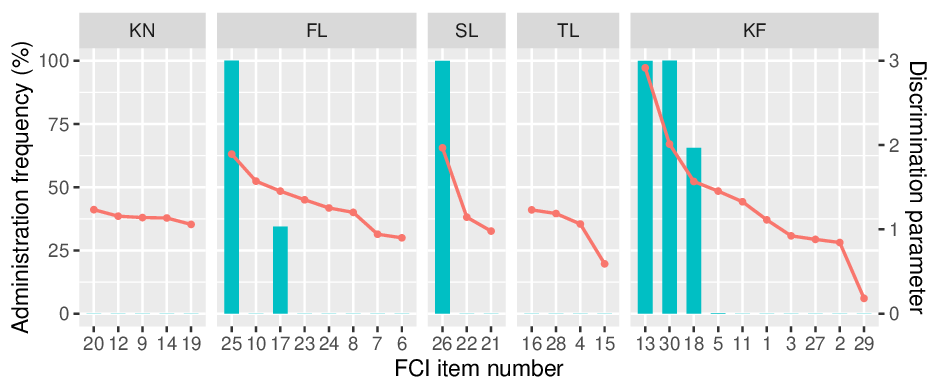} \\
  \includegraphics[width=13cm]{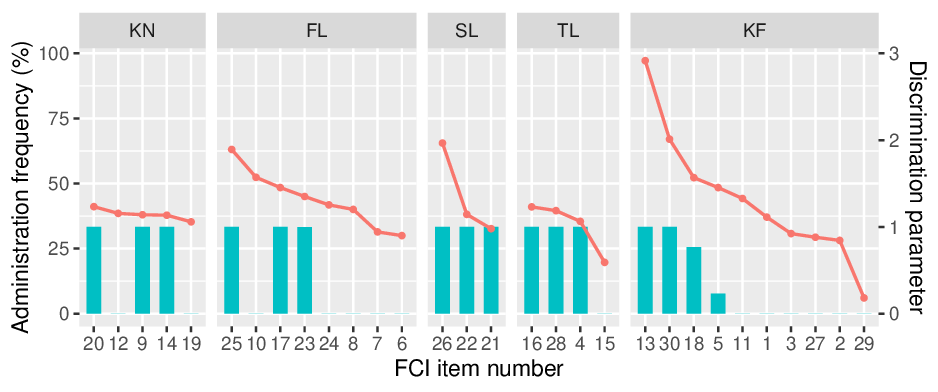} \\
  \end{center}
  \caption{\footnotesize Histogram of the likelihood of an item to be administered on a given test  when there are nine tests of $L=5$ in the Linear model ($0\leq \theta \leq1$).  The red lines show the discrimination parameters for the FCI items.  The graphs are divided by subgroup of the FCI: Kinematics (KN), First law (FL), Second law (SL), Third law (TL), and Kinds of forces (KF).  The top (bottom) graph shows the result without (with) content balancing and item exposure control.}
  \label{Fig4}
  \end{figure*}

Figure~\ref{Fig4} (top) shows histograms of the likelihood for an item to be  administered in each test when there are nine tests of $L=5$ in the Linear model ($0\leq \theta \leq1$).  The graphs are divided into subgroups on the FCI: Kinematics (KN), First law (FL), Second law (SL), Third law (TL), and Kinds of forces (KF).  We can see that some of the items with high discrimination parameters (Q.13, Q25, Q26, and Q30) have a nearly 100\% probability of being administered.  That is, a student would see these four items on each of the nine tests they take.  The fifth item we see would alternate between Q.17 and Q.18.  The other 24 items were very rarely administered to the simulees.

The implementation of content balancing and item exposure control changes this situation, as shown in the right graph of Fig.~\ref{Fig3}.  First, for content balancing, we ensured that the items in each test are not biased toward specific concepts.  For example, since there are five subgroups, there was one item from each subgroup on each test when $L=5$.  In addition to this, following the calculation in Sec.~\ref{subsubsec:algoimp},  we set $N_\textrm{adm}=3$ as the maximum number of times each item could be administered in the nine tests for $L=5$.  We can see that these are realized in the right graph in Fig.~\ref{Fig3}.  However, even with content balancing and item exposure control, there are many cases where an item is never administered in that semester.  This problem is evident in the bottom graph in Fig.~\ref{Fig4}. Compared to the top graph, the number of items with non-zero percentages has increased.  However, nearly half of the items with low discrimination were never administered.  (Although Q.10 and Q.12 have rather high discrimination, they were almost never administered due to their very small difficulty parameters: less than $-1.4$.) One additional issue concerns the variability in difficulty parameters. Since three of the content areas (KN, SL, and TL) contain five or fewer items, the range of difficulty parameters is limited. For example, the difficulty parameters of the SL items are 0.471 for Q.21, $-0.125$ for Q.22, and 0.567 for Q.26. This limited variability may be insufficient to accurately and precisely estimate students' abilities.

These results show that simply expanding the item bank may be insufficient to improve the situation further.  To improve the test efficiency, we suggest creating or incorporating items from other assessment tests \cite{Dancy2006,OsbornPopp2009,Nieminen2010,McCullough2016} that are highly discriminative, i.e. that can better distinguish respondents with different levels of proficiency. In addition, the items should display sufficient variability in their difficulty parameters.

\section{\label{sec:Discussion} Discussion}

\subsection{\label{subsec:summary} Summary}

To address the limitations of the pre-post method, we proposed an alternative approach using a Chain-CAT algorithm that sequentially links the results of each CAT administration in a course using collateral information.  We analyzed the efficiency of the Chain-CAT algorithm through numerical simulations of various progression models for the values of true proficiency.  The simulation results showed that the Chain-CAT algorithm significantly improved the test efficiency in most of the progression models compared to the simple CAT without collateral information.  The simulation results indicated that the accuracy and precision of the Chain-CAT are comparable to that of the pre-post method even when each test length is as short as 5 items when the test is administered 9 times; therefore, the total test length of Chain-CAT can be less than that of the pre-post method.  This finding, however, is without imposing constraints of item balancing and item exposure control.  When these constraints are imposed, we found that the efficiency of the Chain-CAT is lower than that of the pre-post method when the length of each of the nine tests is five items.  These results suggest that the item bank should be expanded by adding new items or incorporating items from other inventories that have high discrimination and exhibit sufficient variability in difficulty parameters.

\subsection{\label{subsec:limitation} Limitations and future work}

\paragraph*{Validating Chain-CAT further}
There are many possibilities for further improving the validity of the Chain-CAT. First, although we found that our FCI dataset has sufficient unidimensionality and local independence at the overall test level \cite{Yasuda2021,Yasuda2022}, higher unidimensionality and local independence are desirable.  This can be realized by dropping one item from each locally dependent pair.  Adding new items would be necessary to compensate for the reduction in efficiency.  Alternatively, using multidimensional models may produce a better fit to the data, resulting in more valid proficiency estimates \cite{Scott2015,Stewart2018,Yang2019,Eaton2020}.   

Second, our content balancing algorithm utilized the FCI taxonomy table \cite{YasudaHull2021}, but researchers have identified issues with this grouping system \cite{Nieminen2010,Huffman1995,Hestenes1995,Heller1995,Scott2012,Semak2017,Eaton2018a}.  When new highly discriminative items (as mentioned above) are added to the item bank, the reliability of the statistical grouping and the content balancing algorithm should be examined.

Third, it is meaningful to conduct Chain-CAT on real classes and assess the validity of the instrument.  We can analyze the validity of the real class data by analyzing the consistency between the Chain-CAT results and other assessment data, such as that obtained in clinical  interviews.  We have conducted a pilot study for validation where we deployed the Chain-CAT system using the mirtCAT package \cite{Chalmers2016} with shinyapps.  Students used their smartphones to complete the Chain-CAT.  The preliminary results showed that Chain-CAT results are roughly consistent with interview findings.  Moreover, the results showed that the progression of proficiency estimates was relatively stable and interpretable—for example, displaying either a steady linear increase or consistently low values—rather than highly fluctuating.  Further work should investigate this more extensively.

Fourth, in Sec.~\ref{subsec:overall}, we calculated the overall average RMSE across various $\sigma^i$ values and based our comparison on the $\sigma^i$ that yielded the lowest overall average RMSE. This approach reflects practical considerations for implementing the test in classroom settings, where instructors determine $\sigma^i$ before administration without knowing the amount of proficiency progression in advance. A possible real-world usage scenario would be one in which the teacher chooses either Chain-CAT (with CI) or the pre-post method. In the former (latter) case, the program would automatically select $\sigma^i$=0.3 ($\sigma^i$=0.6) to minimize RMSE. We found that, even when sigma is optimized, the minimal RMSE when collateral information is not used is larger than the RMSE when collateral information is used.  As such, we do not recommend using Chain-CAT without collateral information. This analysis using session and overall averages may not provide a rigorous comparison, but it can offer a useful reference from a practical perspective that considers the actual conditions of test administration in the classroom. Since our method of analysis is exploratory, there remains room for improvement in such comparative approaches.

Fifth, when expanding the item bank by developing new items, it is important to recognize that evaluating what constitutes a ``good'' question requires multiple perspectives. The results presented in Fig.~\ref{Fig4} illustrate one method of item evaluation, grounded in item response theory. While some items appear to be more discriminating and therefore more informative from the IRT perspective, this approach captures only a portion of what defines item quality. Other important factors—beyond those identified by IRT—may be valued by experts. Thus, to meaningfully assess the quality of new items, additional investigation is needed to determine whether items that perform well under IRT are also regarded as high quality by experts.

\paragraph*{Improving efficiency further}
In addition to the methods we used in this study, there are many options to explore that may further improve the efficiency of the Chain-CAT.  First, while we used the 2PL model, the 3PL model is also applicable. In going from 2PL to 3PL, the overall average RMSE values reported in Table~\ref{table2} were found to improve by approximately 0.01. It is up to the analyst to decide which model to adopt, considering both the degree of improvement and the increased survey burden associated with the larger sample size required for the 3PL model.  Second, although we used the length criterion as the stopping step, an alternative option is to use the precision criterion.  This criterion administers as many items as necessary until pre-specified precision is obtained \cite{Magis2017a}.  
However, because the test length varies from respondent to respondent when the precision criterion is used, it would be inconvenient to administer the test in a classroom setting, so at-home administration may be preferable.  

The algorithm for the use of collateral information could be improved.  In the test step of the Chain-CAT algorithm, we directly used a given student's proficiency estimate; namely, we chose the prior distribution as a normal distribution with $\mu^i=\hat{\theta}_1^i$ for the second test.  The other possibility is predicting the initial proficiency estimate of the $n$th test by a regression model in a similar way to \cite{vanderlinden1999} or another session-series model using the given student's proficiency estimates before the $n$th test.  Another option is to implement the algorithm utilizing the response time in addition to the response choice of the students \cite{vanderlinden2008}.  This information might further improve the test efficiency.

\paragraph*{Possible Practical Uses of Chain-CAT FCI}
One practical application of Chain-CAT is its integration into ordinary short tests administered after university mechanics lectures. For example, each mini test could consist of eight questions: three problems related to the lecture content and five FCI questions presented in CAT format. Students would be graded only on the first three questions, while the latter five would not contribute to their grade. However, because all questions are presented consecutively on the LMS, it is expected that most students will complete the entire test to ensure their evaluation. This design is not only expected to evade declines in response rates but also allows instructors to achieve two complementary goals: The first three problems provide a measure of how well students have understood the lecture's specific content. The five FCI questions assess students' conceptual understanding of Newtonian mechanics in general at the time of the lecture. Throughout the semester, instructors can analyze the changes in conceptual understanding for each student.

As an example of how instructors might interpret Chain-CAT results, we could analyze the impact of specific instructional events, as suggested by Heckler and Sayre \cite{Heckler2010}. For instance, if providing feedback on homework is followed by a sharp increase in students' conceptual understanding, this could serve as evidence of the effectiveness of the instructional method. Furthermore, one may find unexpected changes in the proficiency estimate, for example, decay, as observed in Ref.~\cite{Heckler2010}. The state of understanding that appeared to have temporarily improved was actually unstable and can be interpreted as insufficient understanding. Researchers can use the Chain-CAT as an alternative tool to investigate the nature of student understanding.

Chain-CAT FCI would be particularly suitable for use in first-year university courses with high school physics as a prerequisite. In Japan, most first-year students who study mechanics at the university level have already encountered mechanics in high school, but their high school education tends to emphasize solving problems using formulas. University courses, especially those employing active learning methods such as tutorials, provide a more appropriate context for analyzing how students' conceptual understanding develops.

\paragraph*{Development of Chain-CAT feedback system}
The following issues are needed to develop the feedback system based on Chain-CAT FCI.  First, it is necessary to develop a method to analyze the progression of individual proficiency.  Our simulation results showed a clear progression trend as shown in Fig.~\ref{Fig2}, because the statistics were calculated using 10 000 simulated responses at each session.  However, for an individual respondent, the progression of proficiency fluctuates much more, which makes it difficult to understand the state of the student precisely.  (As mentioned above, although the sample size is small, no significant fluctuations were observed in the surveys conducted with actual students. However, it remains possible that such fluctuations could appear in other contexts.) Therefore, we may need to estimate a student's state based on the analysis of multiple test results, for example, by calculating an average across multiple sessions.  It might be appropriate to group the proficiency estimates into several levels, as in Ref.~\cite{Choi2020}.  

There are many ways to provide feedback from Chain-CAT to respondents \cite{Duss2020}.  A simple score, the proficiency estimate would not be useful for students.  It would be more useful, however, to group the proficiency estimates into levels, for example, simply dividing into four levels by setting thresholds. In Chain-CAT, students can receive feedback in the form of graphs showing their progress, as well as in the form of sentences with encouragement and/or guidance provided by generative artificial intelligence \cite{Kuchemann2024}.  In considering what feedback to provide, student responses not only on the current test but also on prior tests should be considered. For example, a steadily increasing score across tests (indicating that all is well) might be met with automated feedback such as ``I notice your scores are steadily improving. What strategies or habits do you think contributed most to this growth?'' On the other hand, for a student with a stagnant score across several tests (indicating that additional effort is required) feedback might take the form of ``I see your scores haven't changed much yet. What parts of the test felt the hardest for you this time? Which ones felt more manageable?'' In either case, this feedback will promote reflection, metacognition, and self-regulation for the students \cite{Clark2012}. These tasks would be our future work.

\newpage

\appendix*

\section{\label{sec:estimates} Item parameter estimates of the FCI using the 2PL model}

\begin{table}[h]
  \footnotesize
    \centering
    \caption{The results of the estimation for the item parameters of the FCI based on the 2PL model ($N$=2712). The estimates and the standard errors (SEs) for the item parameters are shown.}
  \begin{ruledtabular}
      \begin{tabular}{lrrrrlrrrr} 
   {}  & $\hat{a}$ & SE($\hat{a})$ & $\hat{b}$ &SE($\hat{b})$  &  {}  & $\hat{a}$ & SE($\hat{a})$ & $\hat{b}$ &SE($\hat{b})$ \\  \cline{1-5} \cline{6-10}
   Item 1  & 1.113 & 0.064 & $-$0.961 & 0.059 & Item 16 & 1.231 & 0.066 & $-$0.710 & 0.048 \\
Item 2  & 0.844 & 0.052 & $-$0.425 & 0.056 & Item 17 & 1.453 & 0.071 &  0.525 & 0.041 \\
Item 3  & 0.923 & 0.068 & $-$1.915 & 0.123 & Item 18 & 1.568 & 0.074 & $-$0.069 & 0.036 \\
Item 4  & 1.064 & 0.058 &  0.586 & 0.051 & Item 19 & 1.059 & 0.059 & $-$0.533 & 0.049 \\
Item 5  & 1.453 & 0.070 &  0.225 & 0.038 & Item 20 & 1.233 & 0.063 & $-$0.343 & 0.042 \\
Item 6  & 0.900 & 0.060 & $-$1.365 & 0.088 & Item 21 & 0.980 & 0.055 &  0.471 & 0.052 \\
Item 7  & 0.942 & 0.058 & $-$0.906 & 0.064 & Item 22 & 1.145 & 0.060 & $-$0.125 & 0.043 \\
Item 8  & 1.201 & 0.069 & $-$1.102 & 0.060 & Item 23 & 1.351 & 0.066 &  0.276 & 0.040 \\
Item 9  & 1.140 & 0.061 & $-$0.444 & 0.046 & Item 24 & 1.254 & 0.067 & $-$0.769 & 0.049 \\
Item 10 & 1.571 & 0.096 & $-$1.458 & 0.065 & Item 25 & 1.893 & 0.086 &  0.156 & 0.033 \\
Item 11 & 1.327 & 0.065 &  0.229 & 0.040 & Item 26 & 1.966 & 0.093 &  0.567 & 0.035 \\
Item 12 & 1.156 & 0.079 & $-$1.777 & 0.098 & Item 27 & 0.881 & 0.054 & $-$0.592 & 0.058 \\
Item 13 & 2.916 & 0.139 &  0.061 & 0.028 & Item 28 & 1.188 & 0.064 & $-$0.634 & 0.047 \\
Item 14 & 1.135 & 0.059 &  0.096 & 0.043 & Item 29 & 0.182 & 0.049 & $-$6.645 & 1.799 \\
Item 15 & 0.591 & 0.047 &  0.716 & 0.087 & Item 30 & 2.011 & 0.093 &  0.435 & 0.034 \\
      \end{tabular}%
    \label{table3}%
  \end{ruledtabular}
  \end{table}%

\nocite{*}

\begin{acknowledgments}
This work was supported by JSPS KAKENHI Grant Number JP22H01061 and JP23K22332.
\end{acknowledgments}


%

\end{document}